\documentclass[pra,twocolumn,floatfix,superscriptaddress,amsmath,citeautoscript,aps,longbibliography]{revtex4-1}
\usepackage{graphicx}  % needed for figures
\usepackage{dcolumn}   % needed for some tables
\usepackage{bm}        % for math
\usepackage{amssymb}   % for math
\usepackage{amsmath}
\usepackage{color}
\usepackage{natbib}
\usepackage{hyperref}  % for cite and reference
\hypersetup{hypertex=true,
            colorlinks=true,
            linkcolor=blue,
            anchorcolor=blue,
            citecolor=blue}

\newcommand{\NGG}{Nd$_3$Ga$_5$O$_{12}$}
\newcommand{\GGG}{Gd$_3$Ga$_5$O$_{12}$}

\begin{document}

\title{ Antiferromagnetism and Ising Ground States in the Rare-earth Garnet Nd$_3$Ga$_5$O$_{12}$ }

\date{\today}
\author{N. Zhao}
\thanks{These authors have contributed equally to this work.}
\affiliation{Department of Physics, Southern University of Science and Technology, Shenzhen 518055, China}
\author{H. Ge}
\thanks{These authors have contributed equally to this work.}
\affiliation{Department of Physics, Southern University of Science and Technology, Shenzhen 518055, China}
\author{L. Zhou}
\affiliation{Department of Physics, Southern University of Science and Technology, Shenzhen 518055, China}
\author{Z.~M. Song}
\affiliation{Department of Physics, Southern University of Science and Technology, Shenzhen 518055, China}
\author{J. Yang}
\affiliation{Department of Chemistry, Southern University of Science and Technology, Shenzhen 518055, China}
\author{T.~T. Li}
\affiliation{Department of Physics, Southern University of Science and Technology, Shenzhen 518055, China}
\author{L. Wang}
\affiliation{Department of Physics, Southern University of Science and Technology, Shenzhen 518055, China}
\affiliation{Shenzhen Institute for Quantum Science and Engineering,Shenzhen 518055, China}
\author{Y. Fu}
\affiliation{Department of Physics, Southern University of Science and Technology, Shenzhen 518055, China}
\author{Y.~F. Zhang}
\affiliation{Department of Physics, Southern University of Science and Technology, Shenzhen 518055, China}
\author{J.~B. Xu}
\affiliation{Department of Physics, Southern University of Science and Technology, Shenzhen 518055, China}
\author{S.~M. Wang}
\affiliation{Department of Physics, Southern University of Science and Technology, Shenzhen 518055, China}
\author{J.~W. Mei}
\affiliation{Department of Physics, Southern University of Science and Technology, Shenzhen
  518055, China}
\affiliation{Shenzhen Institute for Quantum Science and Engineering,Shenzhen
  518055, China}
\affiliation{Shenzhen Key Laboratory of Advanced Quantum Functional Materials
   and Devices, Southern University of Science and Technology, Shenzhen 518055, China}
\author{X. Tong}
\affiliation{Institute of High Energy Physics, Chinese Academy of Sciences (CAS), Beijing 100049, China}
\affiliation{Spallation Neutron Source Science Center, Dongguan 523803, China}

\author{L.~S. Wu}
\thanks{Corresponding author: wuls@sustech.edu.cn}
\affiliation{Department of Physics, Southern University of Science and Technology,
   Shenzhen 518055, China}
\affiliation{Shenzhen Key Laboratory of Advanced Quantum Functional Materials
   and Devices, Southern University of Science and Technology, Shenzhen 518055, China}

\author{J.~M. Sheng}
\thanks{Corresponding author: shengjm@sustech.edu.cn}
\affiliation{Department of Physics, Southern University of Science and Technology, Shenzhen 518055, China}
\affiliation{Institute of High Energy Physics, Chinese Academy of Sciences (CAS), Beijing 100049, China}
\affiliation{Spallation Neutron Source Science Center, Dongguan 523803, China}
\affiliation{Academy for Advanced Interdisciplinary Studies, Southern University of Science and Technology, Shenzhen 518055, China}

\date{\today}

\date{\today}

\begin{abstract}
In this paper, we investigate the low temperature magnetic properties of the rare-earth garnet compound \NGG~in detail by means of magnetization, specific heat and magnetocaloric effect measurements. The magnetic thermal properties along with the crystal field calculations reveal that the Nd$^{3+}$ ions form into a frustrated hyper-kagome lattice with connected triangles have an Ising-like ground state with the easy axis along the local [100], [010] and [001] directions. Instead of a quantum spin liquid ground state, an antiferromagnetically ordered state is found below $T_{\mathrm{N}}=0.52~\rm K$. With applying field in the [111] direction, the antiferromagnetic order is suppressed at the critical field of $B_{\mathrm{c}}=0.75~\rm T$, and enhancement of the critical fluctuations with linear crossover behaviors is observed near the critical point.

\end{abstract}

\maketitle

\section{INTRODUCTION}
\maketitle
As an exotic spin state of matter, quantum spin liquid has attracted an extensive amount of interest in condensed matter physics. In comparison to the traditional ferromagnetic or antiferromagnetic spin states, no long-range order that breaks the translational symmetry was expected in the quantum spin liquid state, although the spins are quantum entangled over long distances~\cite{zhouquantum2017,banerjeeneutron2017}. Due to the enormous quantum fluctuations, various exotic spin behaviors were expected in the state of quantum spin liquids. However, the experimental realization of quantum spin liquid state in real materials is a long-sought dream, and many candidate physical models have been proposed. Among these, geometrically frustrated magnetic systems, such as two-dimensional (2D) antiferromagnetic triangle compounds~\cite{rawlba2019,dingpersistent2020,bordelonfieldtunable2019} and kagome lattices~\cite{chenelastic2018,yanspinliquid2011,pereirotopological2014} have been extensively studied. In these 2D spin lattice systems, since the moments are highly frustrated, the entangled spin liquid states with a large degeneracy were expected down to the zero-temperature limit. As an addition, three-dimensional (3D) frustrated spin lattices are very rare, and they have attracted a lot of attention.

Hyper-kagome compounds are unique examples of 3D frustrated spin systems, where magnetic ions form into a 3D network of corner shared triangles. The rare-earth garnet family R$_3$(Al,Ga)$_5$O$_{12}$ is a prototypical example, where the magnetic rare-earth ions reside in two penetrating hyper-kagome lattices of perfect corner-sharing triangles. The most well studied material of this garnet family is \GGG\cite{deen2015,schiffer1994,Ramirez1991,marshall2002,dunsiger2000,schiffer1995,Daudin1982,Kinney1979}, where the magnetic Gd$^{3+}$ ions host an isotropic Heisenberg ground state. Intensive studies of \GGG~have been performed, and peculiar field-temperature phase diagrams were reported~\cite{tsui1999,ghosh2008,hov1980}. Instead of long-range magnetic order, a spin liquid state was proposed~\cite{Petrenko1998,Petrenko1999,TsuiYK2001,Petrenko1997}, and the most recent study indicated that the peculiar magnetic ground states may be raised from the multipolar hidden order~\cite{Joseph2015}. For the garnet systems, large high-quality single crystals are easily accessible, and the most direct way to investigate the spin dynamics would be inelastic neutron scattering. However, the extraordinarily large neutron absorption cross-section of Gd$^{157}$ made neutron scattering experiment on~\GGG~complicated. Alternatively, exploring other rare-earth based garnet systems with smaller neutron absorption cross-sections seems to be a natural option.

In this paper, we chose the Nd-based garnet compound \NGG. In contrast to \GGG, the neutron absorption of Nd isotopes is very weak and this makes future inelastic neutron scattering relatively easier. Different from \GGG~with a classical $S=7/2$ ground state, doublet ground states are realized in \NGG~due to the large crystalline electrical field(CEF), resulting in an effective $S=1/2$ model. Calorimetric study of \NGG~had been carefully performed~\cite{Onn1967}, and it implied that \NGG~entered into an ordered state at $T_{\rm N}=0.516\ \rm K$. Furthermore, it was found that exchange interaction was the dominant factor in the phase transition rather than dipolar interactions~\cite{Onn1967}. Previous neutron diffraction study indicated a ground state with magnetic order described by $\Gamma4$ irreducible representation~\cite{Hammann1968}. Although magnetocaloric effect (MCE) and anisotropy of the magnetic moment at 4.2 K have given us some basic properties of \NGG ~\cite{Onn1967,Nekvasil1974}, detailed studies about the low temperature thermal properties are still needed. Especially, compared to the rich field-induced physics in \GGG, it is still unknown if there is any exotic order that exists in \NGG~when the antiferromagnetic order is suppressed. Here, we perform a comprehensive study of magnetization, heat capacity and MCE down to very low temperatures. An antiferromagnetic ground state has been found. In addition, by applying a magnetic field in the [111] direction, the system is tuned to a critical point around $0.75~\rm T$, and enhanced spin fluctuations are observed.

\section{EXPERIMENTAL DETAILS}

High-quality large single crystals of~\NGG are grown using the Czochralski (CZ) technique, and the crystal structure of \NGG~is confirmed using a Bruker D8 Quest diffractometer with Mo-K$\alpha$ radiation ($\lambda=0.71073~\rm\AA$). All the thermal properties are measured with field along the crystal [111] direction, which is verified using the X-ray Laue camera. Magnetization measurements are performed on the commercial Quantum Design Magnetic Property Measurement System (MPMS), and for temperatures below 1.8 K, a Hall sensor magnetometer integrated with a dilution refrigerator (DR) insert is used~\cite{QDMagnetometry,ACandini2006,ACavallini2004}. Specific heat is measured with the traditional thermal relaxation method, using the commercial Quantum Design Physical Property Measurement System (PPMS). The CEF calculations based on the point charge model were done using the \textsc{McPhase} software~\cite{MCPHASE2004}.

\section{Results and Analysis}

\subsection{Crystal Structure and Crystalline Electrical Field}

\begin{figure}[ht!]
 \includegraphics[width=3.in]{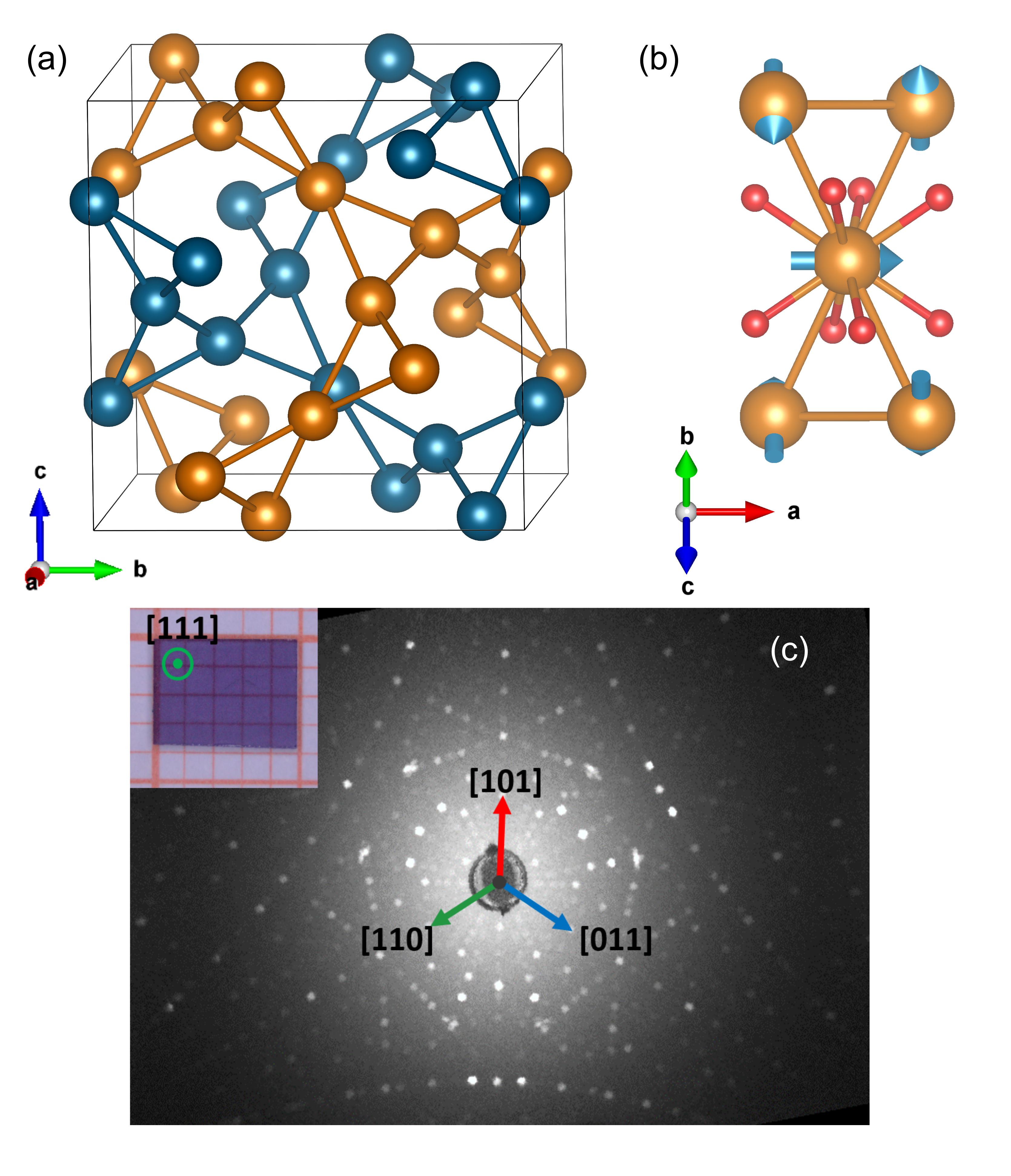}
    \caption{(a) Simplified crystal structure of \NGG, where only the magnetic Nd$^{\rm 3+}$ ions are shown. (b) Calculated moment configurations for Nd$^{\rm 3+}$ magnetic ions (blue arrows), where only eight nearest neighbor O$^{2-}$ charges are considered for the calculations. (c) Measured Laue pattern of the~\NGG~single crystal along the [111] direction. Inset: Single crystal of~\NGG~with the surface normal to the [111] direction. }
\label{structure}
\end{figure}

\NGG~formed in the Garnet crystal structure with the space group \emph{Ia}\={3}\emph{d}. The crystal structure has been verified with single crystal X-ray diffraction. The lattice parameters refined at 100 K were found to be $a=b=c=12.5081{~\rm\AA}$, and $\alpha=\beta=\gamma=90^{\rm\circ}$, which are well consistent with previously reported values~\cite{Sawada1997}. The crystal structure of \NGG~with magnetic Nd$^{\rm 3+}$ sites is shown in Fig.~\ref{structure}(a). In this garnet structure, the magnetic Nd$^{3+}$ ions reside on a 3D network with two interpenetrating corner sharing triangular sublattices. This geometry usually known as hyper-kagome lattice, is equivalent to the kagome lattice in a 2D plane.

In rare-earth-based magnets, CEF configurations are usually important for a comprehensive understanding of the magnetic properties. In the case of Nd$^{3+}$, the good quantum number $J=9/2$ ($L=6, S=3/2$) resulted in ten ($2J+1=10$) multiplet degenerate states, which are lifted into five doublet states due to the environment of the neighboring Oxygen ( O$^{2-}$) and Gallium (Ga$^{3+}$) charges. The energy gaps and the wave functions of these split CEF states are determined by the local point symmetry. For~\NGG,~with the space group \emph{Ia}\={3}\emph{d}, the magnetic Nd$^{3+}$ ions locate at site 24c and have three two-fold axis along the local crystal [011], [0\={1}1]) and [100] directions.
We have performed the CEF calculation based on the point charge model~\cite{Stevens1952,Stevens1952,Hutchings1964,Rotter2011}.
Taking the Nd$^{3+}$ ion located at the position ($0.125,0,0.25$) as an example, the best diagonalization of the CEF Hamiltonian is found when the quantization $z$ axis is chosen along the local [100] direction. The calculated ground state wave functions are:
\begin{multline}\label{E0}
E_{\rm 0\pm} =0.91|\pm 9/2\rangle+0.09|\mp 7/2\rangle+0.39|\pm 5/2\rangle,
\end{multline}
which indicates a well-separated Ising-like ground state with most of the contribution from the $|\pm 9/2\rangle$ states. The calculated saturation moment is found to be $M_{\rm s}\simeq 3~\mu_{\rm B}/\rm Nd$, and the calculated first excited CEF levels are $14.84\rm \ meV\simeq 173\ K$. Applying the above calculations to the other two Nd$^{3+}$ sites at positions (0.25,0.125,0) and (0,0.25,0.125), resulted in the same Ising ground states but with the local easy axis along [010] and [001]. The overall spin configurations of \NGG~are presented in Fig.~\ref{structure}(b), with the Ising axis indicated by the blue arrows. We have to point out here that only the eight nearest neighboring O$^{2-}$ charges are considered in the above calculation, and the calculated CEF configuration is only a rough approximation. These calculated excited energy scheme may not quantitatively match well with the real values, and need to be verified with further neutron scattering experiments.
However, it is the low temperature quantum magnetism that we are interested in here, and the most important information is about the ground states. We found that these CEF calculations suggest an isolated Ising-like ground doublet states with easy axis aligned orthogonal to each other. These results are consistent with previous neutron scattering measurements in zero-field~\cite{Hammann1968}. As indicated by the CEF calculations, the three Nd moments that sharing a triangle are along their local [100], [010] and [001] directions, respectively. Thus, applying magnetic field along [100] direction would be the easy direction for 1/3 of the total Nd$^{3+}$ moments, but the hard direction for the other 2/3 of the total Nd$^{3+}$ moments. Instead, the field direction along [111] was chosen here. With this choice, the angles between the external field and the three local easy moment axis are the same, and thus all the Nd$^{3+}$ moments could be tuned simultaneously in the magnetic field. All the field and temperature-dependent magnetic susceptibility and specific heat data were measured on a single crystal with the large surface normal to [111] axis, and the crystal orientation ~is confirmed by the single crystal Laue camera, as shown in Fig.~\ref{structure}(c).

\subsection{Magnetic Properties}

\begin{figure}[ht!]
 \includegraphics[width=2.8in]{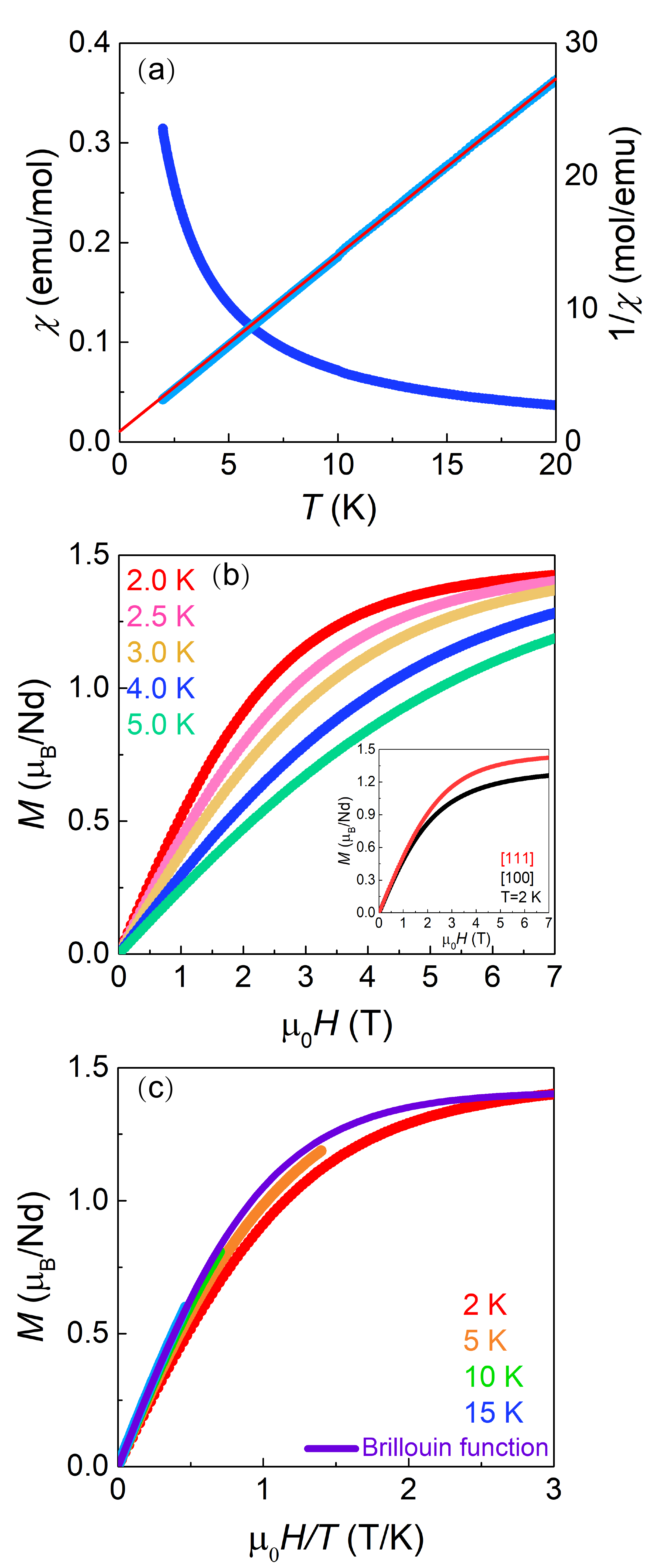}
    \caption{(a) The temperature dependent DC magnetic susceptibility ($\chi\rm$ =$M/H$, left axis) and inverse susceptibility ($1/\chi$, right axis) of \NGG, measured at $\mu_{\rm 0}\rm \textit{H} = 0.1$ T, with the field applied along the crystal [111] direction. The red solid line is the fitting of the Curie-Weiss law. (b) Field dependent DC magnetization measured at different temperatures from 2 K to 5 K, with $\mu_{\rm 0}\rm \textit{H}\parallel \rm [111]$. Inset: Field dependent magnetization with field along the [100] direction. (c) Magnetization measured at different temperatures over-plotted with functions of $\mu_{\rm 0}\rm \textit{H}/\textit{T}$. The calculated Brillouin function assuming an effective $S=1/2$ ground states are presented as the solid purple line.}
\label{magnetization}
\end{figure}

Further characterization of the single crystal magnetization is performed. Fig.~\ref{magnetization}(a) shows the temperature-dependent DC magnetic susceptibility ($\chi\rm$ =$M/H$) of \NGG. The magnetic susceptibility increases as lowering temperatures, and no long-range order is observed down to 2 K. To avoid the contributions from the excited CEF levels at higher temperatures, we fit the inverse magnetic susceptibility with the Curie-Weiss law ($\chi=C/(T-\theta_{\rm cw})$) at a lower temperature range from 2 K to 20 K (red solid line in Fig~\ref{magnetization}(a)). The Curie-Weiss temperature is found to be around $\theta_{\rm cw}=-0.58$~K and the effective moment is found to be $2.45\ \mu_{\rm B}\rm /Nd$. The small negative Curie-Weiss temperature suggests weak antiferromagnetic interactions presented at low temperatures.

The field-dependent magnetization measured at different temperatures from 2 K to 5 K is shown in Fig.~\ref{magnetization}(b). The saturation moment achieved at 2 K and 7 T is about $1.42\ \mu_{\rm B}\rm /Nd$. As discussed above, the CEF calculations suggest that the Nd$^{3+}$ moments are pointing along the three orthogonal [100], [010] and [001] directions, and neither of them is fully along [111] direction in the saturation field. Taking the angle $\varphi=55^{\rm o}$ between these local easy axis and the [111] field direction into consideration, the expected saturation moment should be $M_{\rm s}(\mu_{\rm 0}\rm \textit{H}\parallel111)$=$M_{\rm s}$cos$\varphi\simeq1.71\ \mu_{\rm B}\rm /Nd$, that is close to the experimentally observed value $1.42 \ \mu_{\rm B}\rm /Nd$. The field-dependent magnetization with field along [100] direction is shown in the inset of Fig.~\ref{magnetization}(b). The magnetization moment is about $1.2\ \mu_{\rm B}\rm /Nd$ at 2 K and 7 T. This value is slightly smaller than the saturation moment with field along [111] direction, since the other 2/3 of the Nd$^{3+}$ moments are not fully polarized along their local [100] direction. However, for the future study, it will be interesting to measure the magnetization with field along [100] up to high magnetic fields, where all the Nd$^{3+}$ moments would be fully polarized.

Magnetization as functions of $\mu_{\rm 0}\rm \textit{H}/\textit{T}$ measured at different temperatures are plotted in Fig.~\ref{magnetization}(c). The calculated Brillouin function assumes a doublet ground state with effective spin $S=1/2$, and $g_{\rm eff}=2.84$ is over-plotted (the solid purple line). The magnetization curves at higher temperatures around 10 K and 15 K are well consistent with the calculated Brillouin function, however, when the measuring temperatures are comparable to the Curie-Weiss temperature $\theta_{\rm cw}=-0.58~\rm K$, the measured data indicates significant deviations from the Brillouin function curve, due to the antiferromagnetic interactions.

\subsection{Zero-field Specific Heat}
\begin{figure}[ht!]
 \includegraphics[width=2.8in]{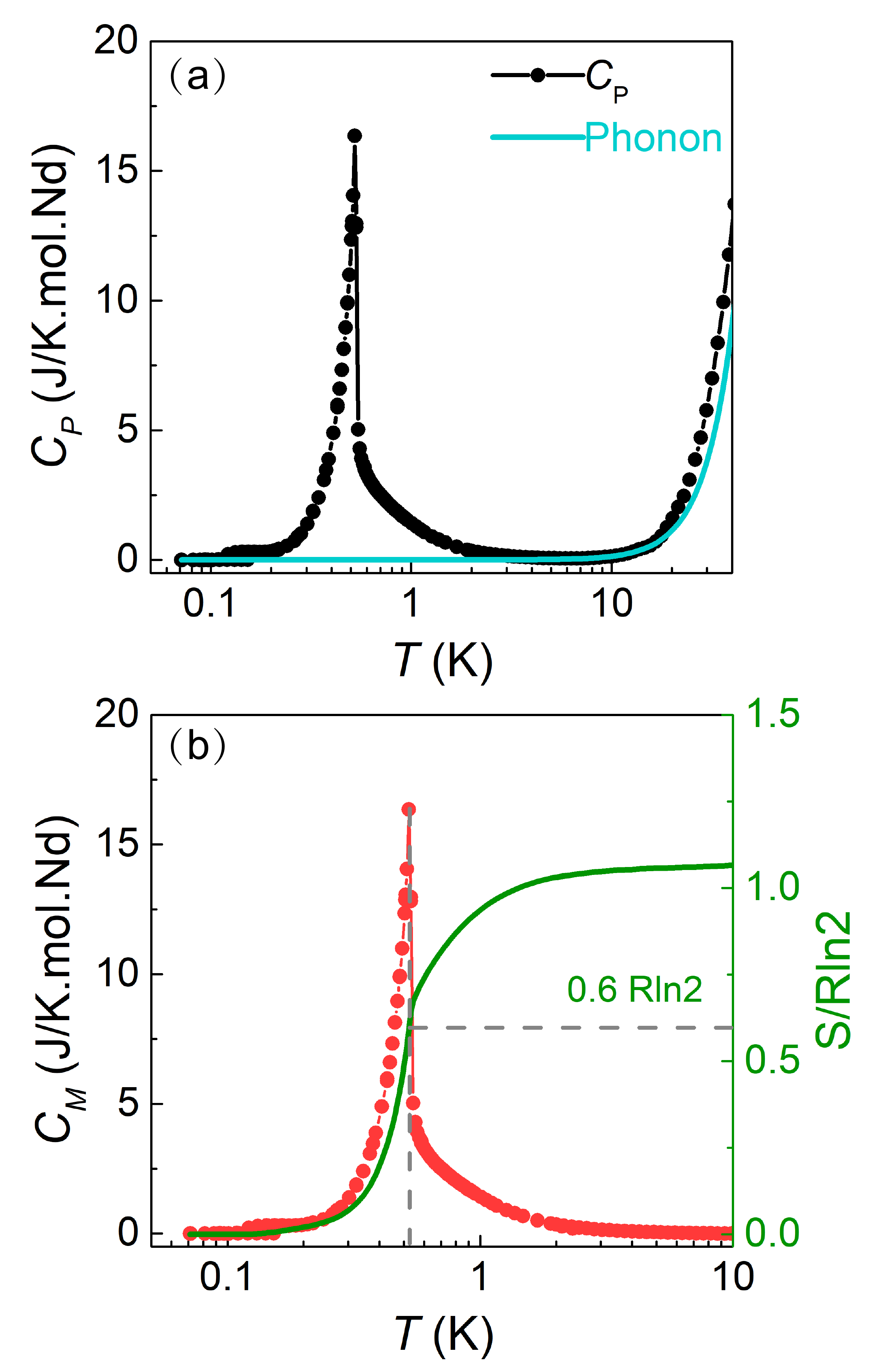}
    \caption{(a) Temperature dependent specific heat ($C_{\rm P}$) measured at zero-field (black solid circles), and the phonon contribution estimated from the Debye model (blue line). (b) Magnetic specific heat ($C_{\rm M}$) with phonon contributions corrected (red solid circles), and the integrated magnetic entropy (green solid line). The long range antiferromagnetic order is indicated by the gray dashed lines.}
\label{HC}
\end{figure}

To explore the low temperature properties of \NGG, specific heat measurements are performed. Fig.~\ref{HC}(a) shows the zero-field heat capacity, $C_{\rm P}$, down to 50 mK. Higher temperature specific heat is fitted with the Debye model and the magnetic specific heat with phonon contribution corrected is shown in Fig.~\ref{HC}(b). At the transition temperature $T_{\rm N}=0.52\ \rm K$, a sharp peak-like anomaly is observed, followed by a long tail that extended to temperatures as high as 2 K. The integrated entropy is presented in Fig.~\ref{HC}(b). A full entropy of $R$ln2 is achieved above 2 K. It is interesting to notice that about $60\% R\ln2$ of the total entropy is released at the antiferromagnetic transition, while the rest is left at higher temperatures above the phase transition.
\subsection{Field-Temperature Phase Diagram}
\begin{figure}[ht!]
 \includegraphics[width=2.7in]{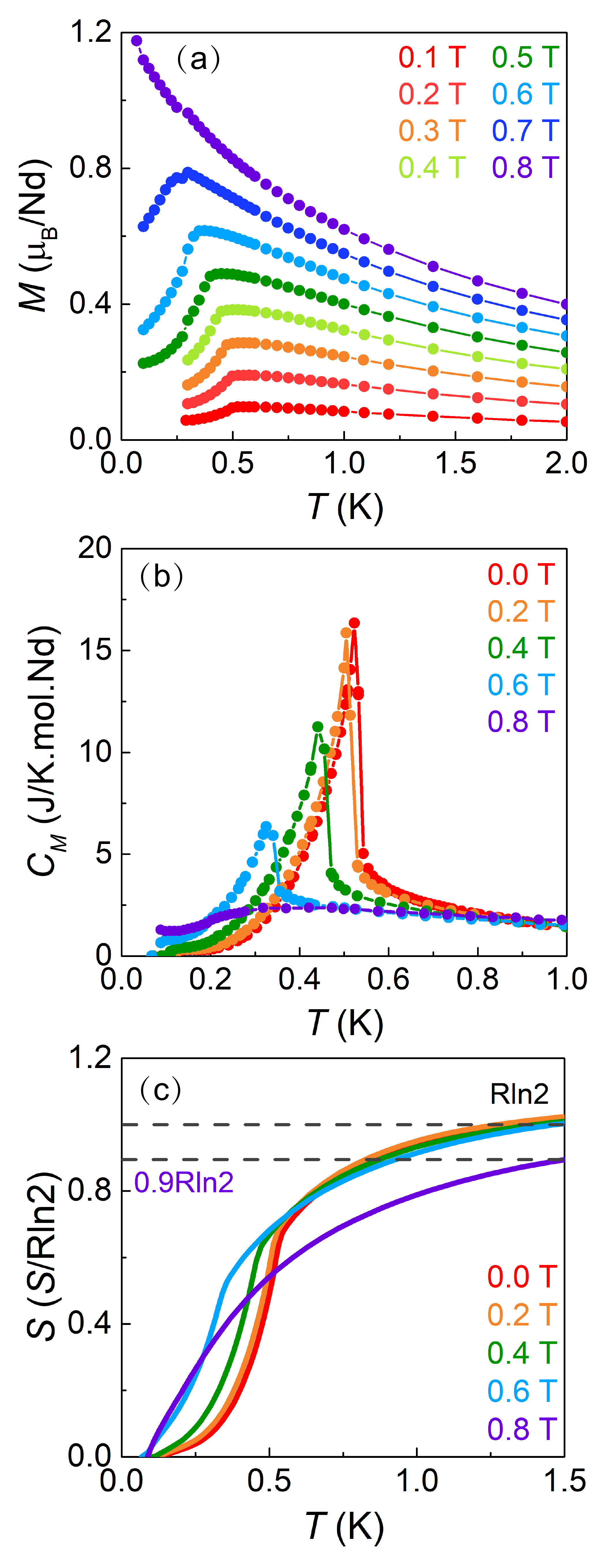}
    \caption{(a)-(b) The temperature dependent magnetization (a) and magnetic specific heat (b) measured under different magnetic field along the [111] direction. (c) The temperature dependence of the integrated entropy at different magnetic fields. }
\label{MT}
\end{figure}

To further investigate the magnetically ordered phase, magnetization measurements down to 0.1 K are performed with field along the [111] direction. As shown in Fig.~\ref{MT}(a), a cusp-like feature is observed at the transition temperature. This cusp-like feature is then gradually suppressed with increasing fields. Finally, the antiferromagnetic order disappeared above the critical field around 0.8 T. Similar behaviors are also observed in the specific heat measurements, as presented in Fig.~\ref{MT}(b). The sharp triangular-like peak is suppressed to lower temperatures with increasing fields and finally turned into a broad anomaly around 0.8 T. The integrated magnetic entropy at different fields is presented in Fig.~\ref{MT}(c). In zero-field, the obtained magnetic entropy at the $T_{\rm N}$ is about $60\%R$ln2. With increasing fields, only about $44\%R$ln2 is reached at $T_{\rm N}$ for $\mu_{\rm 0}\rm \textit{H}=0.6$ T. However, in all these field ranges from 0 to 0.6 T, a full expected $R$ln2 is obtained at temperatures above 1.4 K. In contrast, the integrated entropy at the critical field 0.8 T only reached about $90\%R$ln2. Since the experimental data is not integrated from the exact zero-temperature, the contributions below 0.1 K are ignored. Normally little fluctuations are left in zero-temperature. However, at the critical point, quantum fluctuations are playing an important role, and the missing $10\%R$ln2 of entropy at 0.8 T comes from the residual entropy enhanced by the quantum fluctuations.

\begin{figure}[ht!]
 \includegraphics[width=2.8in]{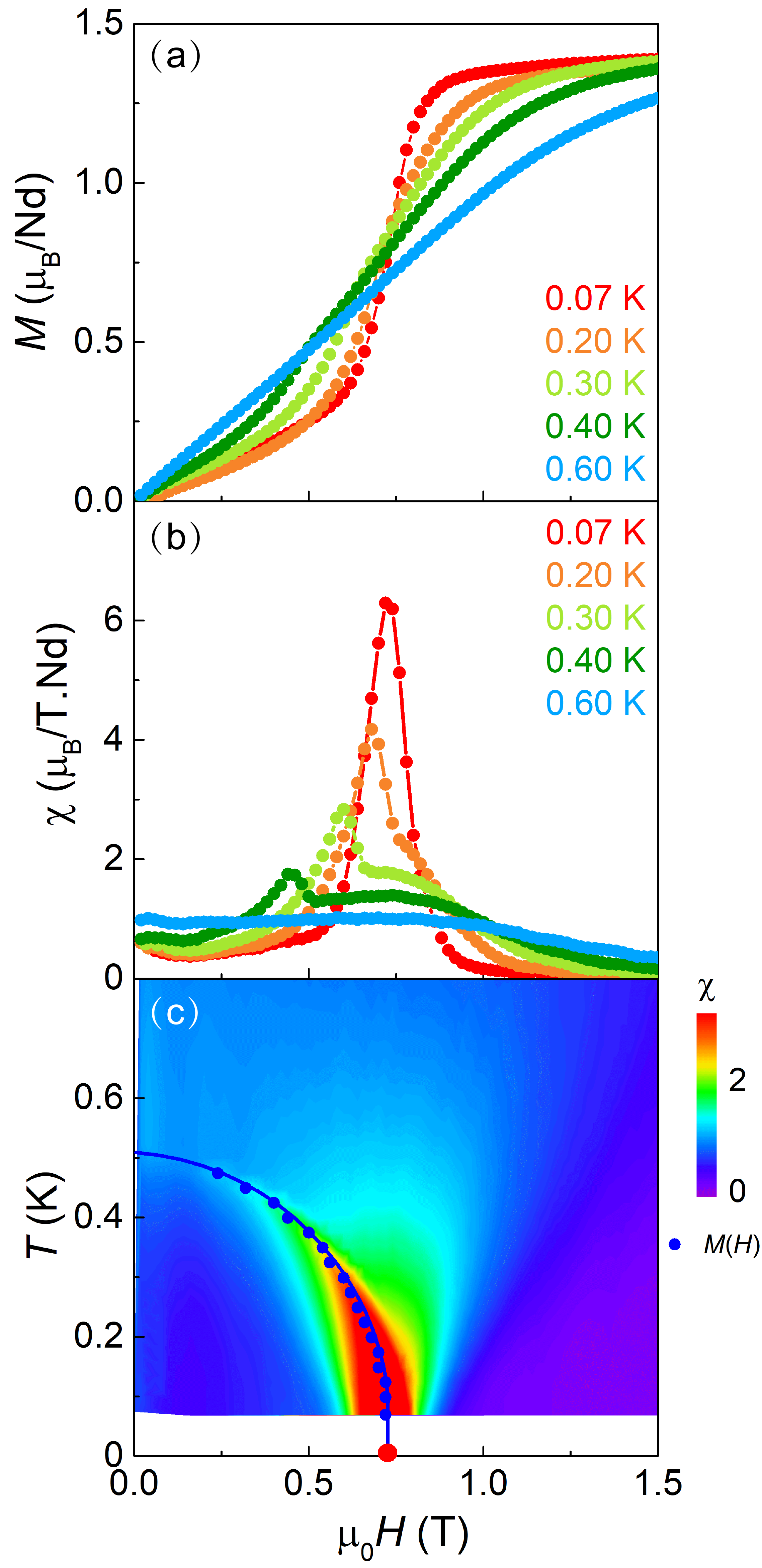}
    \caption{(a)-(b) Field dependent magnetization $M$ (a) and magnetic susceptibility $\chi=dM/\mu_{\rm 0}\rm \textit{dH}$ (b) measured at different temperatures from 0.6 K down to 0.07 K. (c) Field temperature ($\mu_{\rm 0}\rm \textit{H}$-\textit{T}) phase diagram of \NGG.}
\label{MH}
\end{figure}

The field-dependent magnetization below the ordering temperature is shown in Fig.~\ref{MH}(a). As for lowering temperature, a steep increase in magnetization is observed at the critical field. The transition fields are defined by the peak positions in magnetic susceptibility, as shown in Fig.~\ref{MH}(b). At intermediate temperatures, such as 0.4 K and 0.3 K, broad anomalies are always observed followed by the peak-like features. Finally at the base temperature around 0.07 K, these two features merged into a broad peak with width at half maximum around 0.14 T. The overall field-temperature phase diagram with the phase boundaries extracted from the temperature and field-dependent magnetization is plotted in Fig.~\ref{MH}(c)(blue solid circles ). Together with the phase boundary, the contour of the field-dependent susceptibility at different temperatures is over-plotted in Fig.~\ref{MH}(c) as well. As discussed above, there is a broad region with a width about $0.15\sim0.2$ T centered at 0.75 T, indicating a great number of fluctuations persist near the critical point.

\begin{figure}[ht!]
 \includegraphics[width=2.8in]{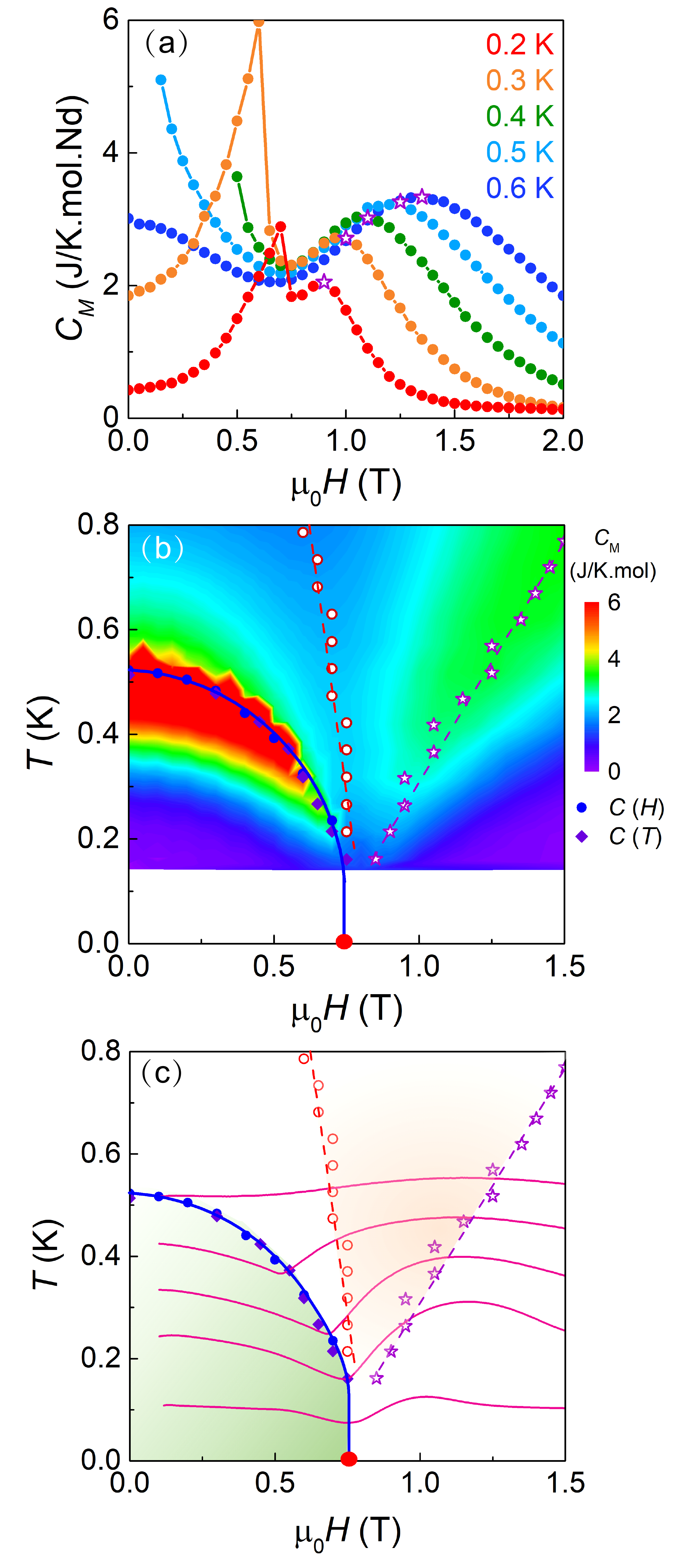}
    \caption{(a) The field-dependent specific heat $C_{\rm M}$ of \NGG~at different temperatures. (b) Contour plot of the field-dependent specific heat, and the field temperature phase diagram. The linear crossovers extracted from the peak positions of the field-dependent specific heat are indicated by the empty stars and the dash lines. (c) The measured MCE of \NGG~at different temperatures. }
\label{MCE}
\end{figure}

Similar to the field-dependent magnetization, the field-dependent specific heat is presented in Fig.~\ref{MCE}(a). A sharp decrease is observed at the phase boundary, followed by a minimum at the critical field. More interestingly, another broader anomaly is observed above the critical field. The maximum positions of the peak increased linearly with increasing temperatures, as indicated by the empty stars and the blue line in Fig.~\ref{MCE}(b). This effect is not due to the Schottky effect in magnetic fields. In the case of a Schottky anomaly of a two-level system, the maximum of the specific heat would stay at a constant value around 3.6 J/mol.K for different fields. However, in Fig.~\ref{MCE}(a), the maximum value above the critical fields is significantly smaller than 3.6 J/mol.K at the base temperatures. These observed behaviors are the crossovers in the vicinity of a gapless quantum critical point (QCP). This is consistent with the observations of the MCE measurements. As shown in Fig.~\ref{MCE}(c), the red solid lines are the measured temperature-field dependence as sweeping magnetic field in a quasi-adiabatic condition. Clear dip-like features are observed at the phase boundary at different temperatures. As keeping increasing fields, a broad peak-like anomaly is observed near the crossover line, just as reported in other quantum magnets~\cite{Nicolo2018,Coldea2001}.

\section{Conclusion}

In summary, we have performed a comprehensive investigation of the specific heat, magnetization of the single crystal \NGG. The CEF calculations based on the point charge model reveal that the ground doublet states of the magnetic Nd$^{3+}$ ions are Ising-like, with the local easy axis along [100], [010] and [001] directions. The magnetic thermal properties indicate an antiferromagnetic ground state in zero-field below $T_{\rm N}=0.52$ K. This antiferromagnetic state is tuned with magnetic field, resulting in a critical point around 0.75 T. The specific heat and magnetic entropy indicate that not all the magnetic moments of Nd$^{3+}$ ions are well ordered and part of the local moments are keeping fluctuating at the AFM ground states. Especially, as the system is tuned by the magnetic field to the vicinity of the quantum critical point, enhancement of critical fluctuations and linear crossover behaviors are observed.

However, considering that the magnetic Nd$^{3+}$ ions are located in a geometrically frustrated hyper-kagome lattice with corner-sharing triangles, quantum ground states are expected. In contrast to the isostructure compound \GGG, where a great amount of fluctuations and series of complex quantum states in fields have been reported~\cite{deen2015,hov1980,schiffer1994,Ramirez1991,marshall2002,dunsiger2000,tsui1999,ghosh2008,TsuiYK2001,Petrenko1997,Joseph2015,
schiffer1995,Daudin1982,Kinney1979,Petrenko1998,Petrenko1999}, the physics here in \NGG~ appears to be rather classical. A significant difference between these two compounds is that the Gd$^{3+}$ ions host almost ideal Heisenberg ground states of $S=7/2$, due to the absence of crystal field, while the Nd$^{3+}$ magnetic moments are forced to align orthogonally to each other along their local Ising axis. This situation is very similar to the other rare-earth-based garnet family compounds, such as Er$_3$Ga$_5$O$_{12}$~\cite{Cai2019}, Dy$_3$Al$_5$O$_{12}$~\cite{Keen1966,Steiner1982,Joshua1981}, Tb$_3$Al$_5$O$_{12}$~\cite{Joshua1981} and Ho$_3$Ga$_5$O$_{12}$~\cite{Zhou2008}. We conclude that the strong single ion anisotropy arises from the CEF effect has lifted the extensive degeneracy, and thus release a great number of fluctuations. In addition, the CEF analysis indicates three orthogonally aligned Nd$^{3+}$ moments. With this moment configuration, the term $\mathbf{M}_{1}\cdot \mathbf{M}_{2}$ of the nearest neighbor dipole interaction would vanish. This may be the reason why the dipole interaction is weak, although the nearest neighbor distance is not that large.  This is consistent with the conclusion found in the previous report~\cite{Onn1967} that the exchange interaction was the dominant factor in the zero-field antiferromagnetic phase transition of~\NGG. On the other hand, significant critical fluctuations are observed near the field-induced critical point, and further inelastic neutron scattering around that quantum critical region would be rather interesting.

\begin{acknowledgments}
The research at SUSTech was supported by the National Natural Science Foundation of China (No.~12134020 ). Part of this work was also supported by the National Natural Science Foundation of China (No.~11974157), (No.~11875265) and (No.~12104255). The Scientific Instrument Developing Project of the Chinese Academy of Sciences ($^{3}$He based neutron polarization devices), the Institute of High Energy Physics, the program for Guangdong Introducing Innovative and Entrepreneurial Teams (No.~2017ZT07C062), by Shenzhen Key Laboratory of Advanced Quantum Functional Materials and Devices (No.~ZDSYS20190902092905285). The authors acknowledge the assistance of SUSTech Core Research Facilities.

\end{acknowledgments}


\begin{thebibliography}{10}

\bibitem{zhouquantum2017}
Y. Zhou, K. Kanoda, and T. K. Ng, Quantum spin liquid states, Rev. Mod. Phys., 89, 025003 (2017).

\bibitem{banerjeeneutron2017}
A. Banerjee, J. Yan, J. Knolle, and C. A. Bridges, Neutron scattering in the proximate quantum spin liquid $\alpha-$RuCl$_{3}$, Science, 356, 1055 (2017).

\bibitem{rawlba2019}
R. Rawl, L. Ge, Z. Lu, Z. Evenson, C. R. Dela Cruz, Q. Huang, M. Lee, E. S. Choi, M. Mourigal, H. D. Zhou, and J. Ma, Ba$_{8}$MnNb$_{6}$O$_{24}$:A model two-dimensional spin-5/2 triangular lattice antiferromagnet, Phys. Rev. Mater., 3, 054412 (2019).

\bibitem{dingpersistent2020}
Z. Ding, Z. Zhu, J. Zhang, C. Tan, Y. Yang, D. E. MacLaughlin, and L. Shu, Persistent spin dynamics and absence of spin freezing in the H-T phase diagram of the two-dimensional triangular antiferromagnet YbMgGaO$_{4}$, Phys. Rev. B, 102, 014428 (2020).

\bibitem{bordelonfieldtunable2019}
M. M. Bordelon, E. Kenney, C. Liu, T. Hogan, L. Posthuma, M. Kavand, Y. Lyu, M. Sherwin, N. P. Butch, C. Brown, M. J. Graf, L. Balents, and S. D. Wilson, Field-tunable quantum disordered ground state in the triangular-lattice antiferromagnet NaYbO$_{2}$, Nat. Phys., 15, 1058 (2019).

\bibitem{chenelastic2018}
H. Chen, H. Nassar, A. N. Norris, G. K. Hu,and G. L. Huang, Elastic quantum spin Hall effect in kagome lattices, Phys. Rev. B, 98, 094302 (2018).

\bibitem{yanspinliquid2011}
S. Yan, D. A. Huse, and S. R. White, Spin-liquid ground state of the S=1/2 kagome Heisenberg antiferromagnet, Science, 332, 1173 (2011).

\bibitem{pereirotopological2014}
M. Pereiro, D. Yudin, J. Chico, C. Etz, O. Eriksson, and A. Bergman, Topological excitations in a kagome magnet, Nat. Commun., 5, 1 (2014).


\bibitem{deen2015}
P. P. Deen, O. Florea, E. Lhotel, and H. Jacobsen, Updating the phase diagram of the archetypal frustrated magnet Gd$_{3}$Ga$_{5}$O$_{12}$,  Phys. Rev. B, 91, 014419 (2015).

\bibitem{schiffer1994}
P. Schiffer, A. P. Ramirez, D. A. Huse, and A. J. Valentino, Investigation of the field induced antiferromagnetic phase transition in the frustrated magnet: Gadolinium Gallium Garnet, Phys. Rev. Lett., 73, 2500 (1994).

\bibitem{Ramirez1991}
A. P. Ramirez, and R. N. Kleiman, Low‐temperature specific heat and thermal expansion in the frustrated garnet Gd$_{3}$Ga$_{5}$O$_{12}$,  J. Appl. Phys., 69, 5252 (1991).

\bibitem{marshall2002}
I. M. Marshall, S. J. Blundell, F. L. Pratt, A. Husmann, C. A. Steer, A. I. Coldea, W. Hayes1 and R. C. C. Ward, A muon-spin relaxation ($\mu$SR) study of the geometrically frustrated magnets Gd$_{3}$Ga$_{5}$O$_{12}$ and ZnCr$_{2}$O$_{4}$, J. Phys.: Condens. Matter, 14, L157 (2002).

\bibitem{dunsiger2000}
S. R. Dunsiger, J. S. Gardner, J. A. Chakhalian, A. L. Cornelius, M. Jaime, R. F. Kiefl, R. Movshovich, W. A. MacFarlane, R. I. Miller, J. E. Sonier, and B. D. Gaulin, Low temperature spin dynamics of the geometrically frustrated antiferromagnetic garnet Gd$_{3}$Ga$_{5}$O$_{12}$, Phys. Rev. Lett., 85, 3504 (2000).

\bibitem{schiffer1995}
P. Schiffer, A. P. Ramirez, D. A. Huse, P. L. Gammel, U. Yaron, D. J. Bishop, and A. J. Valentino, Frustration induced spin freezing in a site-ordered magnet: gadolinium gallium garnet. Phys. Rev. Lett., 74, 2379 (1995).

\bibitem{Daudin1982}
B. Daudin, R. Lagnier, and B. Salce, Thermodynamic properties of the gadolinium gallium garnet, Gd$_{3}$Ga$_{5}$O$_{12}$, between 0.05 and 25 K, J. Magn. Magn. Mater., 27, 315 (1982).

\bibitem{Kinney1979}
W. I. Kinney, and  W. P. Wolf, Magnetic interactions and short range order in gadolinium gallium garnet. J. Appl. Phys., 50, 2115 (1979).

\bibitem{tsui1999}
Y. K. Tsui, C. A. Burns, J. Snyder, and P. Schiffer, Magnetic field induced transitions from spin glass to liquid to long range order in a 3D geometrically frustrated magnet, Phys. Rev. Lett., 82, 3532 (1999).

\bibitem{ghosh2008}
S. Ghosh, T. F. Rosenbaum, and G. Aeppli, Macroscopic signature of protected spins in a dense frustrated magnet, Phys. Rev. Lett., 101, 157205 (2008).


\bibitem{hov1980}
S. Hov, H. Bratsberg, and A. T. Skjeltorp, Magnetic phase diagram of gadolinium gallium garnet, J. Magn. Magn. Mater., 15, 455 (1980).

\bibitem{Petrenko1998}
O. A. Petrenko, C. Ritter, M. Yethiraj, and D. M. Paul, Investigation of the low-temperature spin-liquid behavior of the frustrated magnet gadolinium gallium garnet. Phys. Rev. Lett., 80, 4570 (1998).

\bibitem{Petrenko1999}
O. A. Petrenko, D. M. Paul, C. Ritter, T. Zeiske, and M. Yethiraj, Magnetic frustration and order in gadolinium gallium garnet. Physica B: Condensed Matter, 266, 41 (1999).

\bibitem{TsuiYK2001}
Y. K. Tsui, J. Snyder, and P. Schiffer, Thermodynamic study of excitations in a three-dimensional spin liquid, Phys. Rev. B, 64, 012412 (2001).

\bibitem{Petrenko1997}
O. A. Petrenko, C. Ritter, M. Yethiraj, and D. M. K. Paul, Spin-liquid behavior of the gadolinium gallium garnet, Physica B, 241, 727 (1997).

\bibitem{Joseph2015}
J. A. M. Paddison, H. Jacobsen, O. A. Petrenko, M. T. Fernández-Díaz, P. P. Deen, and A. L. Goodwin, Hidden order in spin-liquid Gd$_{3}$Ga$_{5}$O$_{12}$, Science, 350, 179 (2015).

\bibitem{Onn1967}
D. G. Onn, H. Meyer, and J. P. Remeika. Calorimetric study of several rare-earth gallium garnets. Phys. Rev., 156, 663 (1967).

\bibitem{Hammann1968}
J. Hammann, Etude par diffraction de neutrons a 0.31 K de la structure magnetique du grenat de neodyme et de gallium, Phys. Lett. A, 26, 263 (1968).

\bibitem{Nekvasil1974}
V. Nekvasil,V. Roskovec, F. Zounova, and P. Novotný. Anisotropy of the magnetic moment of neodymium in garnets. Czech. J. Phys. B, 24, 810 (1974).

\bibitem{QDMagnetometry}
Magnetometry by means of Hall micro-probes in the Quantum Design PPMS, Quantum Design Application Note, 1084-701.

\bibitem{ACavallini2004}
A. Cavallini, Deep levels in MBE grown AlGaAs$/$GaAs heterostructures, Microelectron. Eng., 73, 954 (2004).

\bibitem{ACandini2006}
A. Candini, G. C. Gazzadi, A. di Bona, M. Affronte, D. Ercolani, G. Biasiol, and L. Sorba, Hall nano-probes fabricated by focused ion beam, Nanotechnology, 17, 2105 (2006).

\bibitem{MCPHASE2004}
M. Rotter, Using McPhase to calculate magnetic phase diagrams of rare earth compounds, J. Magn. Magn. Mater., 272, E481 (2004).

\bibitem{Sawada1997}
H. Sawada, Electron density study of garnets: Z$_{3}$Ga$_{5}$O$_{12}$;Z=Nd, Sm, Gd, Tb, J. Solid State Chem., 132, 300 (1997).


\bibitem{Stevens1952}
K. W. H. Stevens, Matrix elements and operator equivalents connected with the magnetic properties of rare earth ions, Proc. Phys. Soc. A, 65, 209 (1952).

\bibitem{Hutchings1964}
M. T. Hutchings, Point-charge calculations of energy levels of magnetic ions in crystalline electric fields, Solid State Phys., 16, 227 (1964).

\bibitem{Rotter2011}
M. A. Rotter, B. Fåk, and J. A. Blanco, Magnetic excitations in the longitudinally amplitude modulated magnetic structure of PrNi$_{2}$Si$_{2}$, J. Phys. Conf. Ser. 325, 012008 (2011).


\bibitem{Nicolo2018}
N. Defenu, T. Enss, M. Kastner, and G. Morigi, Dynamical critical scaling of long-range interacting quantum magnets, Phys. Rev. Lett., 121, 240403 (2018).


\bibitem{Coldea2001}
R. Coldea, D. A. Tennant, A. M. Tsvelik, and Z. Tylczynski, Experimental realization of a 2D fractional quantum spin liquid, Phys. Rev. Lett., 86, 1335 (2001).

\bibitem{Cai2019}
Y. Cai, M. N. Wilson, J. Beare, C. Lygouras, G. Thomas, D. R. Yahne, K. Ross, K. M. Taddei, G. Sala, H. A. Dabkowska, A. A. Aczel, and G. M. Luke, Crystal fields and magnetic structure of the Ising antiferromagnet Er$_{3}$Ga$_{5}$O$_{12}$, Phys. Rev. B, 100, 184415 (2019).

\bibitem{Keen1966}
B. E. Keen, D. Landau, B. Schneider, and W. P. Wolf, First‐and higher‐order magnetic phase transitions in dysprosium aluminum garnet , J. Appl. Phys., 37, 1120 (1966).

\bibitem{Steiner1982}
M. Steiner and N. Giordano, Neutron-diffraction study of the magnetic structure of dysprosium aluminum garnet, Phys. Rev. B, 25, 6886 (1982).

\bibitem{Joshua1981}
J. Felsteiner and S. K. Misra, Low-temperature ordered states of dysprosium, terbium, and holmium aluminum garnets, Phys. Rev. B, 24, 2627 (1981).

\bibitem{Zhou2008}
H. D. Zhou, C. R. Wiebe, L. Balicas, Y. J. Yo, Y. Qiu, J. R. D. Copley, and J. S. Gardner, Intrinsic spin-disordered ground state of the Ising garnet Ho$_{3}$Ga$_{5}$O$_{12}$, Phys. Rev. B, 78, 140406 (2008).



\end{thebibliography}
\end{document}